\documentclass[a4paper]{article}
\usepackage{newtxtext} 
\usepackage{newtxmath}
\usepackage[T1]{fontenc}
\usepackage{multicol}
\usepackage{titlesec}
\usepackage{enumerate}
\usepackage{enumitem}
\usepackage{booktabs}
\usepackage{array}
\usepackage{hyperref}
\usepackage{fancyhdr}
\usepackage{tabularx}
\usepackage{lipsum}
\usepackage{graphicx}
\usepackage{acronym}

    \acrodef{DER}{Distributed Energy Resource}
    \acrodef{DSO}{Distribution System Operator}
    \acrodef{LV}{Low-Voltage}
    \acrodef{MV}{Medium-Voltage}
    \acrodef{LVN}{Low-Voltage Network}
    \acrodef{OPF}{Optimal Power Flow}
    \acrodef{LP}{Linear Program}
    \acrodef{MINLP}{Mixed Integer Non-Linear Program}
    \acrodef{MIQCP}{Mixed Integer Quadratically Constrained Program}
    \acrodef{MILP}{Mixed Integer Linear Program}
    \acrodef{ToU}{Time-of-Use}
    \acrodef{RTP}{Real-Time Pricing}
    \acrodef{DSOC}{Distribution System Operator Controlled}
    \acrodef{NCL}{Non-Controllable Load}
    \acrodef{HP}{Heat Pump}
    \acrodef{COP}{Coefficient of Performance}
    \acrodef{PV}{Photovoltaic}
    \acrodef{EV}{Electric Vehicle}
    \acrodef{BESS}{Battery Energy Storage System}
    \acrodef{SOC}{State of Charge}
    \acrodef{HEMS}{Home Energy Management System}
    \acrodef{OA}{Outer Approximation}

\hypersetup{
    colorlinks=true, 
    linkcolor=red,   
    urlcolor=blue,   
    citecolor=green,  
    pdfborderstyle={/S/U/W 1} 
}
\setlist[itemize]{leftmargin=*}

\titleformat{\subsection}      
  {\itshape}                      
  {\thesubsection.}               
  {0.5em}                         
  {}                              
\titlespacing*{\subsection}{0pt}{1ex}{1em} 

\titleformat{\subsubsection}[runin]
  {\itshape}                      
  {\thesubsubsection.}            
  {0.5em}                         
  {}
\titlespacing*{\subsubsection}{0pt}{1.5ex plus .1ex minus .2ex}{1.5ex plus .2ex}

\usepackage[
    top=1in,
    bottom=1in,
    left=0.59in,
    right=0.59in,
    headheight=61pt, 
    headsep=0.2in, 
    footskip=0.49in
]{geometry}

\makeatletter
\def\ciredmaketitle{
    \vspace*{-1 cm} 
    
    \begin{center}
        {\fontsize{18pt}{22pt}\selectfont \bfseries \@title \par} 
        \vspace{0.3cm} 
        
        {\lineskip 0.5em 
        {\fontsize{12pt}{14pt}\selectfont \bfseries \itshape \@author \par}
        }
        
        \vspace{-.9 cm} 
        
        \@date \par
    \end{center}
    \par
    \vspace*{0.5cm} 
}
\makeatother

\usepackage[numbers]{natbib}
\usepackage{algorithm,algorithmic}
\usepackage{caption}
\usepackage{float}
\usepackage{xcolor}

\allowdisplaybreaks

\begin{document}

\title{PROSUMER SYNCHRONISATION RISK: IMPACTS OF TIME-VARYING TARIFFS ON DISTRIBUTION NETWORK EXPANSION}
\author{Dillon Zadoks$^{\dagger,1}$, Alfredo Oneto$^{\dagger,2}$$^*$, Carlo Tajoli$^1$, Yi Guo$^{3,4}$, Philipp Heer$^3$, Giovanni Sansavini$^2$, Gabriela Hug$^1$} 
\date{} 


\ciredmaketitle 
 \begin{center}

    $^1$Power Systems Laboratory, Institute for Power Systems \& High Voltage Technology, Department of Information Technology and Electrical Engineering, ETH Z\"urich, Zurich, Switzerland.\\
    $^2$Reliability and Risk Engineering Laboratory, Institute of Energy and Process Engineering, Department of Mechanical and Process Engineering, ETH Z\"urich, Zurich, Switzerland.\\
    $^3$Urban Energy Systems Laboratory, Swiss Federal Laboratories for Materials Science and Technology, Z\"urich, Switzerland.\\
    $^4$Beijing Institute of Technology, Beijing, China.\\
    $\dagger$ These authors contributed equally to this work.\\
    $^*$aalfredo@ethz.ch%

 \end{center}
\thispagestyle{fancy}
 \begin{center}
  \end{center}
\begin{multicols}{2}
\section*{Abstract}
The rapid deployment of distributed energy resources, including heat pumps, electric vehicles, photovoltaics, and battery storage, is reshaping the operation of low-voltage networks. While distribution system operators often aim to develop time-varying tariffs to incentivise network-friendly behaviour and defer reinforcements, they risk triggering prosumer synchronisation. As prosumers follow price signals, their behaviour may synchronise, avoiding existing load peaks while creating undesirable peaks at other times. This paper quantifies the impact on low-voltage network reinforcement needs when shifting from flat tariffs to two time-varying tariffs: Time-of-Use and Real-Time Pricing. Applying a mixed-integer linear model across 471 low-voltage reference networks in Switzerland, we evaluate expansion requirements under projections of distributed energy resource deployment for 2050. In the model, prosumers operate their distributed energy resources in accordance with their subscribed electricity tariff to minimise costs, and the distribution system operator subsequently optimises reinforcements. Our results demonstrate that while adopting time-varying tariffs can reduce reinforcement investment costs by more than 50\%, high adoption rates can reverse this declining trend. Specifically, once prosumer adoption exceeds the optimal thresholds of 50\% for Time-of-Use and 40\% for Real-Time Pricing, reinforcement needs increase by 39\% and 134\%, respectively, for an additional 20\% of tariff adoption.
\section{Introduction} \label{sec:intro}


The increasing electrification of space heating and transportation sectors~\cite{iea2025global}, together with widespread adoption of \ac{PV} and \acp{BESS}~\cite{irena2025renewable}, is accentuating the stress on \acp{LVN}~\cite{KOTSONIAS2025111258}. To prevent voltage and current violations under these conditions, significant network reinforcements may be required. While widespread electrification introduces new challenges for \acp{LVN}, energy accumulation and power shifting are possible with flexible electricity demand, such as \acp{HP}, \acp{EV}, and \acp{BESS}. Therefore, if they are appropriately managed, they can alleviate network stress, potentially reducing reinforcement needs~\cite{navidi2023der}.

One promising approach is to adopt network tariffs to actively encourage network-friendly behaviour. 
Time-varying electricity tariffs, such as \ac{ToU}~\cite{filippiniShortlongrun2011} and \ac{RTP}~\cite{HU2025125815}, are already widely implemented, featuring higher and lower cost periods throughout the day. Moreover, the increasing adoption of price-responsive home management systems and smart devices allows \acp{DSO} to increase the time granularity of tariffs, consequently improving their ability to alleviate network congestion. However, providing the same tariff signal to all devices in the network may have drawbacks. If a large number of devices are aware of upcoming prices and optimise their operations to reduce costs, they may synchronise their operations, causing rebound effects and potentially new load peaks that lead to voltage or current violations. Providing different tariffs to different customers may also be deemed unfair and entail the risk of bias on both sides.


Existing studies have examined the future deployment of \acp{DER} and their impact on \ac{LVN} operation and reinforcements using heuristic methods to identify reinforcement plans or to report reductions in equipment violations~\cite{GUPTA2021116504, anya1}. These methods enable large-scale case studies but may yield suboptimal planning decisions due to their heuristic algorithms. In~\cite{navidi2023der}, a centralised algorithm is used to optimally plan the operation of \acp{DER}, showing significant decreases in operational limit violations at the cost of increased electricity procurement expenses. Additionally, studies investigating time-varying tariffs to shape \ac{DER} operational behaviour often assume full network adoption and perfect price sensitivity~\cite{anya1}. Nevertheless, in practice, prosumers have varying willingness to adopt time-varying tariffs and are not perfectly price-responsive.

In this work, we propose a \ac{MILP} model to investigate optimal reinforcement plans for \acp{LVN}, considering nodes with \acp{DER} that minimise their costs under different time-varying tariffs. We analyse the effects of these tariffs on the network reinforcement at varying levels of tariff adoption. The case study is conducted on 471 \ac{LVN} models using future projections of \ac{DER} installations.

Our contributions are summarised as follows:  
\begin{enumerate}
    \item We propose an optimisation model for network reinforcement planning and \ac{DER} operations to enable efficient investment decisions.
    \item  We analyse the impact of time-varying tariffs on reinforcement needs across different levels of adoption. The model assumes that a fraction of the flexible resources actively optimise their operations to minimise costs, while the others are price-agnostic.
    \item We assess the performance of tariffs across 471 \acp{LVN} with projected \acp{DER} deployments for 2050. The results demonstrate that time-varying tariff adoption of up to 50\% can reduce reinforcement costs by more than 50\%, while high adoption rates can reverse this declining trend.
\end{enumerate} 

In the rest of this paper, we describe the modelling of the flexible devices, the \acp{LVN}, and the optimisation problems in Section~\ref{sec:method}, we introduce the case study in Section~\ref{sec:cs}, we present and discuss the results in Section~\ref{sec:experiments}, and Section~\ref{sec:conclusions} concludes the paper.


\section{Methodology}
\label{sec:method}

The methodology section is divided into three parts. First, we present the \acp{DER} and their modelling (Section~\ref{sec:dermodel}). Second, we introduce the electricity tariffs considered in this work (Section~\ref{sec:tariffs}). Third, we formulate a tractable distribution network expansion planning model (Section~\ref{sec:expansion}).

We represent vectors with bold symbols; $p$ and $q$ denote active and reactive power, respectively. Indices $i$ and $j$ represent nodes; $t$, $d$, and $s$ index the time steps, days, and scenarios, respectively. Upper and lower bounds for each variable are defined as $\overline{\cdot}$ and $\underline{\cdot}$, respectively.

\subsection{Distributed Energy Resources} \label{sec:dermodel}

The following constraints are applied to the \acp{DER} for each node and scenario. The indices are omitted to improve readability.

PV systems are modelled assuming no active power curtailment to promote renewable power generation. The constraint is as follows:
\begin{equation}
    \boldsymbol{p}^{\mathrm{PV}} = \boldsymbol{p}^{\mathrm{PV}}_{\mathrm{gen}} ,\label{eq:pvgenconstr}
\end{equation}
\noindent fixing \ac{PV} active power output. 

\ac{BESS} co-installations with \ac{PV} systems are considered in the model. The storage dynamics of \acp{BESS} are given by:
\begin{equation}
    \mathrm{SOC}_{t+1} = \mathrm{SOC}_{t} + \eta^{\mathrm{BESS}^+} p_{t}^{\mathrm{BESS}^+} - \frac{1}{\eta^{\mathrm{BESS}^-}} p_{t}^{\mathrm{BESS}^-} \forall\, t,
    \label{eq:bessbal}
\end{equation}
\noindent which relates the \ac{SOC} at the next time step to that at the current time step, where $\eta^{\mathrm{BESS}^+},\eta^{\mathrm{BESS}^-}$ denote charging/discharging efficiencies. For cyclic storage modelling, the last \ac{SOC} is set equal to the initial \ac{SOC}. In addition, we consider \ac{SOC} and power limits, as well as linear constraints to reduce simultaneous charging and discharging~\cite{Pozo_2022}.

The thermal dynamics of buildings with installed \acp{HP} are modelled using a resistor-capacitor representation, with temperature $T$ as a state variable and thermal parameters $C$ and $H$~\cite{zapparoli}:
\begin{subequations}\label{eq:hps}
\begin{align}
C\bigl(T_{t+1} - T_{t}\bigr)
&= \mathrm{COP}_{t}\, p^{\mathrm{HP}}_t -H\bigl(T_t- T_t^{\text{amb}}\bigr) \quad \forall t, \label{eq:hpbal}\\
\underline{\boldsymbol{T}} &\leq \boldsymbol{T} \leq \overline{\boldsymbol{T}}, \quad \underline{T}^{\mathrm{avg}} \leq T^{\mathrm{avg}}.\label{eq:tconstr}
\end{align}
\end{subequations}

Constraint~\eqref{eq:hpbal} relates the indoor temperature of the next step to the current one, considering a time-dependent \ac{COP} for the \ac{HP} and positive heating power within allowable capacity limits. Constraints~\eqref{eq:tconstr} bound the temperature within the thermal comfort range, and the average temperature is lower bounded to avoid an overly optimistic operation that tracks the minimum allowable temperature.

Unidirectional smart charging \ac{EV} modelling is adopted from~\cite{herrera2025modelingchargingdemandquantifying}, which considers realistic \ac{EV} fleet baseline charging profiles and their flexibility. The constraints are as follows:
\begin{subequations}\label{eq:evs}
\begin{align}
\underline{\boldsymbol{p}}^{\text{EV}} &\leq \boldsymbol{p}^{\text{EV}} \leq \overline{\boldsymbol{p}}^{\text{EV}} \label{eq:ev_power_bounds} \\
 \boldsymbol{\delta}^{\text{EV}} &\geq \left|\boldsymbol{p}^{\text{EV,base}} - \boldsymbol{p}^{\text{EV}} \right| \label{eq:ev_delta} \\
 \sum_{t \in \mathcal{T}_d} \delta_t^{\text{EV}} &\leq F_d \qquad \forall d  \label{eq:ev_p_sum} \\
  \sum_{d} \sum_{t \in \mathcal{T}_d} p_t^{\text{EV}}& = \sum_{d} \sum_{t \in \mathcal{T}_d} p_t^{\text{EV,base}},  \label{eq:ev_total} 
\end{align} \label{eq:evconstraints}
\end{subequations}

\noindent with the profile shifting denoted by $\delta_t^{\text{EV}}$, and the daily charging flexibility limit defined by $F_d$. The \ac{EV} charging demand is bounded between the upper and lower charging profiles in~\eqref{eq:ev_power_bounds}. Constraint~\eqref{eq:ev_delta} tracks the absolute deviation of the charging profile from its base profile. In addition,~\eqref{eq:ev_p_sum} limits the total energy that the controlled profile can shift during each operation day, and~\eqref{eq:ev_total} ensures that the charged energy during a week is equal to the baseline.

\subsection{Electricity Tariffs} \label{sec:tariffs}

Electricity tariffs are generally composed of three parts: electricity supply cost, network usage cost, and electricity taxes. Furthermore, their design can be static or time-varying. We consider flat tariffs, \ac{ToU} tariffs, and \ac{RTP} tariffs, illustrated in Figure~\ref{fig:tariffs} and explained in the following. We assume prosumers subscribe to a specific tariff and optimally schedule their flexible \acp{DER} to minimise electricity costs.

\begin{center} 
    \captionsetup{hypcap=false}
    \includegraphics[width=0.48\textwidth]{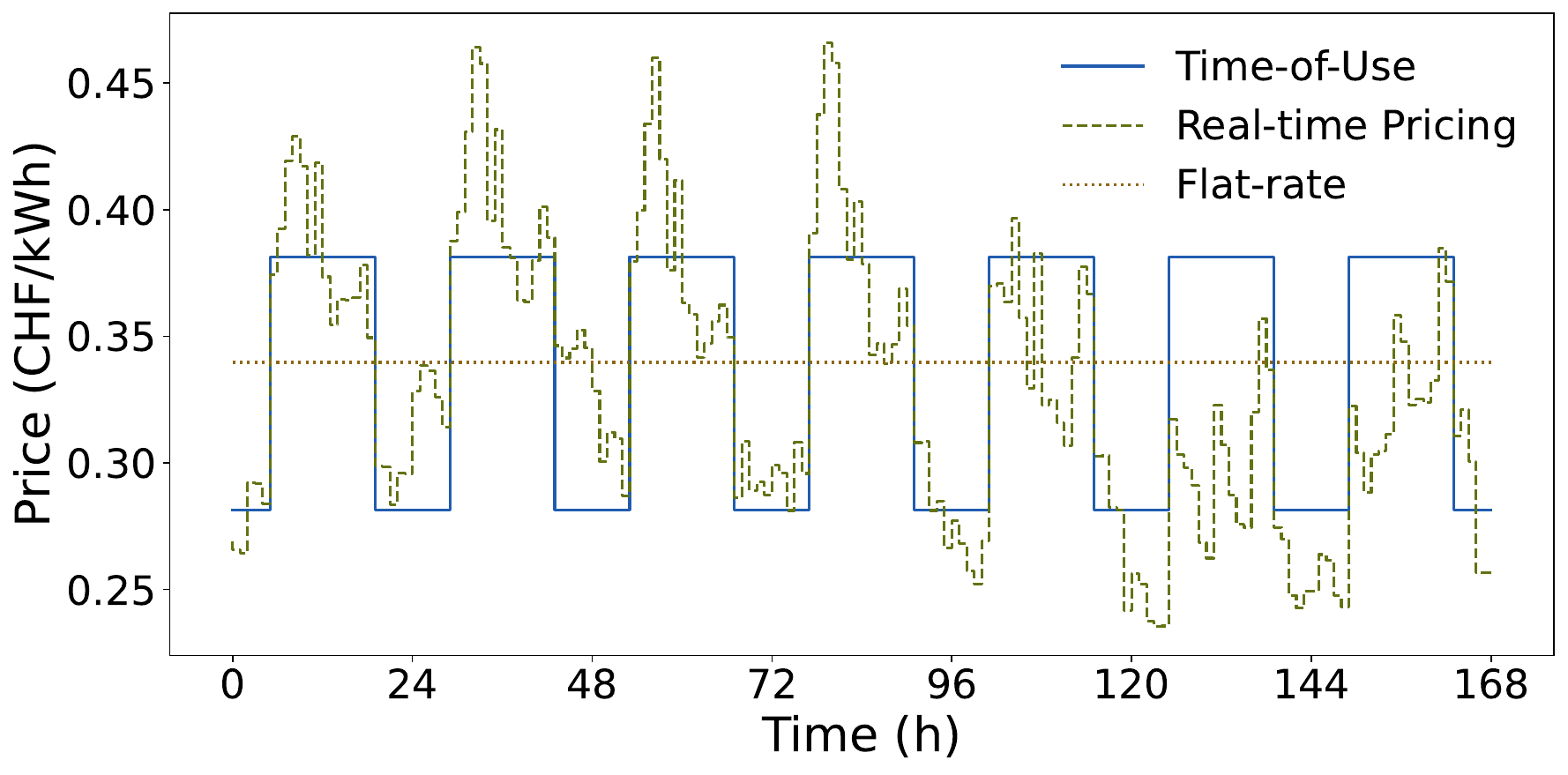}
    \captionof{figure}{Comparison of Flat, \ac{ToU}, and \ac{RTP} tariff structures, based on the case study data.}
    \label{fig:tariffs}
\end{center}

\subsubsection{Flat Tariff:} \label{sec:flat}

This tariff considers a single, uniform price that remains constant throughout the day and year.

\subsubsection{Time-of-Use Tariff:} \label{sec:tou}

Under this scheme, the \ac{DSO} publishes different tariff rates for predefined periods, typically distinguishing between peak and off-peak hours. This structure encourages prosumers with \acp{DER} to shift demand from high-cost peak periods to low-cost off-peak periods and maximise self-consumption during high-priced periods.

\subsubsection{Congestion-Based Real-Time Pricing Tariff:} \label{sec:rtp}

The \ac{RTP} tariff is an approach designed to reflect the actual cost of network usage. To do so, the \ac{DSO} publishes day-ahead or week-ahead prices based on forecasted network loading. In this work, we use the forecasted transformer loading as a proxy for network congestion and formulate the \ac{RTP} tariff as follows:
\begin{equation}
  \mathrm{RTP}(t) = B \cdot \left(\alpha + \beta
  \frac{L(t)}{L^{\mathrm{avg}}} + \gamma\right), \label{eq:rtp}
\end{equation}
\noindent where $B$ is the flat tariff, $\alpha$ is the electricity supply component, $\beta$ is the network usage component, and $\gamma$ is the tax component. $L(t)$ denotes the forecasted transformer loading at time $t$, while $L^{\mathrm{avg}}$ represents the average transformer loading over the time horizon. By construction and choice of $\alpha$, $\beta$, and $\gamma$, the average electricity tariff over the period remains equal to the flat tariff.

\subsection{Distribution Network Expansion} \label{sec:expansion}

Since prosumers are unaware of network constraints, they make decisions on the operation of their \acp{DER} solely based on the \ac{DSO}'s electricity prices and their consumption requirements. Hence, we determine prosumer consumption by minimising their electricity costs while enforcing power balances without restricting flows. Based on the obtained power flows, the \ac{DSO} minimises reinforcement costs through expansion planning.

The prosumer's optimisation, for given tariff and feed-in vectors ($\boldsymbol{c}^{\text{tariff}}$ and $\boldsymbol{c}^{\text{feed-in}}$) is as follows:
\begin{subequations}\label{eq:ldf}
  \begin{align}
  & \min_{\boldsymbol{p}, \boldsymbol{q}} \quad (\boldsymbol{c}^{\text{tariff}})^{\top} \boldsymbol{p}^{e} - (\boldsymbol{c}^{\text{feed-in}})^{\top} \boldsymbol{p}^ g\\
   & \text{s.t.} \quad
\boldsymbol{p}^{l}_{i,j} + \boldsymbol{p}^g_{j} = \boldsymbol{p}^e_{j} + \sum_{n \in K(j)} \boldsymbol{p}^{l}_{j,n}
    && \forall \, (i,j) 
    \label{eq:active_power_bal} \\
    &\boldsymbol{q}^{l}_{i,j} + \boldsymbol{q}^g_{j} = \boldsymbol{q}^e_{j} + \sum_{n \in K(j)} \boldsymbol{q}^{l}_{j,n}
    && \forall \, (i,j) 
    \label{eq:reactive_power_bal} \\
    & \eqref{eq:pvgenconstr}\text{--}\eqref{eq:evconstraints},\label{eq:ders}
  \end{align} \label{eq:deroptimization}
\end{subequations}


\noindent with superscripts $e$, $g$, and $l$ representing the active and reactive power demand, generation, and line flows. Equations~\eqref{eq:active_power_bal} and~\eqref{eq:reactive_power_bal} impose the power balance at the network nodes, considering a linearised radial branch flow, where $K(j)$ denotes the set of nodes connected downstream from node $j$.

The varying level of adoption is accounted for in problem~\eqref{eq:deroptimization} by assigning a fraction of the nodes' \acp{DER} to a time-varying tariff. At the same time, we assign the remaining prosumers to the flat tariff, with the tariff vector adjusted accordingly. Once the prosumers have taken their decisions, we retrieve the active and reactive power flows from the optimal solution of problem~\eqref{eq:deroptimization}, namely $(\boldsymbol{p}^l)^{*}$ and $(\boldsymbol{q}^l)^{*}$ for each line. Then, the \ac{DSO} minimises transformer and line expansions subject to different operational scenarios:
\begin{subequations}
  \begin{align}
    &\min_{\boldsymbol{z}^{\text{trafo}}, \boldsymbol{z}^{\text{lines}}} \quad f^{\text{trafo}} z^{\text{trafo}}+ \sum_{i,j} f^{\text{line}}_{i,j} z_{i,j}^\text{line} \label{eq:objexpansion}\\
    &\text{s.t.} \quad \boldsymbol{V}_{0} = \boldsymbol{V}_{\text{slack}} \label{eq:ss_slack}\\
    &\boldsymbol{v}_{i,s} = \boldsymbol{V}_{i,s}^2 && \forall \, i, s \label{eq:volt_square}\\
    &\underline{\boldsymbol{V}} \leq \boldsymbol{V}_{i,s} \leq \overline{\boldsymbol{V}}
    && \forall \, i, s
    \label{eq:voltageconstr} \\
    &\boldsymbol{v}_{i,s} - \boldsymbol{v}_{j,s} = 2\frac{r_{i,j} (\boldsymbol{p}_{i,j}^{l})_s^{*} + x_{i,j} (\boldsymbol{q}_{i,j}^{l})_s^{*}}{1 + z^{\text{line}}_{i,j}}
    && \forall \, (i,j), s
    \label{eq:volt_drop} \\
   &\left \| \begin{pmatrix}
    (p_{i,j,t}^{l})_s^{*}\\
    (q_{i,j,t}^{l})_s^{*}
    \end{pmatrix} \right \|_2\leq \overline{S_{i,j}^{\text{line}}} \left(1 +
    z_{i,j}^{\text{line}} \right)
    && \forall (i,j), t, s\label{eq:ampacityconstr}\\
   &\left \| \begin{pmatrix}
    p_{t,s}^{\text{trafo}}\\
    q_{t,s}^{\text{trafo}}
    \end{pmatrix} \right \|_2\leq \overline{S^{\text{trafo}}} \left(1 + z^{\text{trafo}} \right)
    && \forall \, t, s.\label{eq:trafoconstr}
  \end{align} \label{eq:distexpansion}
\end{subequations}

The objective function ~\eqref{eq:objexpansion} minimises the integer line and transformer expansions ($\boldsymbol{z}^{\text{lines}}$ and $\boldsymbol{z}^{\text{trafo}}$), with respective costs of $\boldsymbol{f}^{\text{line}}$ and $\boldsymbol{f}^{\text{trafo}}$. 
Equation~\eqref{eq:ss_slack} sets the slack bus at node 0, where the transformer is located;~\eqref{eq:volt_square}--\eqref{eq:volt_drop} indicate the squared voltage, the voltage magnitude limits, and the linearised branch flow equation~\cite{Yeh2012}, with its adapted version with line expansions~\cite{Wei}. Moreover,~\eqref{eq:ampacityconstr} and~\eqref{eq:trafoconstr} indicate the line and transformer ampacity constraints. The power balance constraints are omitted for compactness, although they follow the same structure as the balance equations in~\eqref{eq:deroptimization}. The investment decision variables are scenario-independent and must be robust for all scenarios. Constraints~\eqref{eq:volt_square}--\eqref{eq:trafoconstr} are scenario-dependent, and are applied for the realization of $(\boldsymbol{p}^l)_s^{*}$ and $(\boldsymbol{q}^l)_s^{*}$ in each scenario $s$. 

To reduce the computational complexity, we linearise the non-linear constraints~\eqref{eq:ampacityconstr}--\eqref{eq:trafoconstr} via polyhedral outer approximations. Furthermore, we equivalently linearise the voltage drop constraint in~\eqref{eq:volt_drop} using McCormick envelopes with binary representations of the integer variables~\cite{Wei}.

\section{Case Study}
\label{sec:cs}

In this Section, we present the case study setup, which analyses \acp{LVN} with \ac{DER} projections in the city of Bern, Switzerland.  

\subsection{Network Models and Distributed Energy Resources} \label{sec:networkdata}

We use open \ac{LVN} models covering the considered region, operating at 400~V~\cite{onetopdgs}. These data are derived from modelled, georeferenced medium- and low-voltage networks in Switzerland. The specifications of projected \acp{PV}, co-located \acp{BESS}, \acp{EV}, and non-controllable loads in 2050 are retrieved from the Swiss database published by~\cite{zapparoli}. Moreover, the \ac{HP} allocations and thermal modelling parameters are provided by~\cite{guo2025}.

\subsection{Economic Parameters} \label{sec:economic}

We adopt standard prices for \ac{LVN} equipment investments~\cite{consentec2006costmodelseng, dena_DistributionNetworkStudy_2012}, considering that the reference \acp{LVN} in the region have 630 kVA transformers and NAVY4x240SE cables. The costs, expressed in thousands of CHF, are 15 per km of cables, 85 per km of construction work, and 20 per transformer. In addition, for capital investment annualisation, the lifespan of cables and transformers is assumed to be 40 and 20 years, respectively. 

To construct the electricity tariffs, we use consumption and feed-in prices published by the Bern \ac{DSO}~\cite{EWB_homepage_2025}. The total flat supply cost, in Raps (CHF cents) per kWh, is 33.97, composed of 14.59 for power consumption, 14.03 for network usage, and 5.35 for tax, while the feed-in compensation is 14.80. As the \ac{DSO} does not offer a \ac{ToU} tariff, we construct one with a 10~Raps/kWh difference between the on-peak period (06:00 to 20:00) and the off-peak price. The \ac{RTP} tariff is calculated using~\eqref{eq:rtp} with the reported coefficients. The transformer loading $L(t)$ is estimated by calculating the power flows with no DER flexibility usage.

\subsection{Operational Scenarios} \label{sec:opuncertainty}

For each climatic season, we segment the profiles into 13-week periods. We compute the mean vectors and covariance matrices of the weekly segments independently for electricity prices, temperature, nodal demand, and aggregated \ac{PV} generation, ensuring inter-network consistency. We then generate 20 weekly profiles for each \ac{LVN} using Gaussian sampling.


\begin{center} 
    \captionsetup{hypcap=false}
    \includegraphics[width=0.48\textwidth]{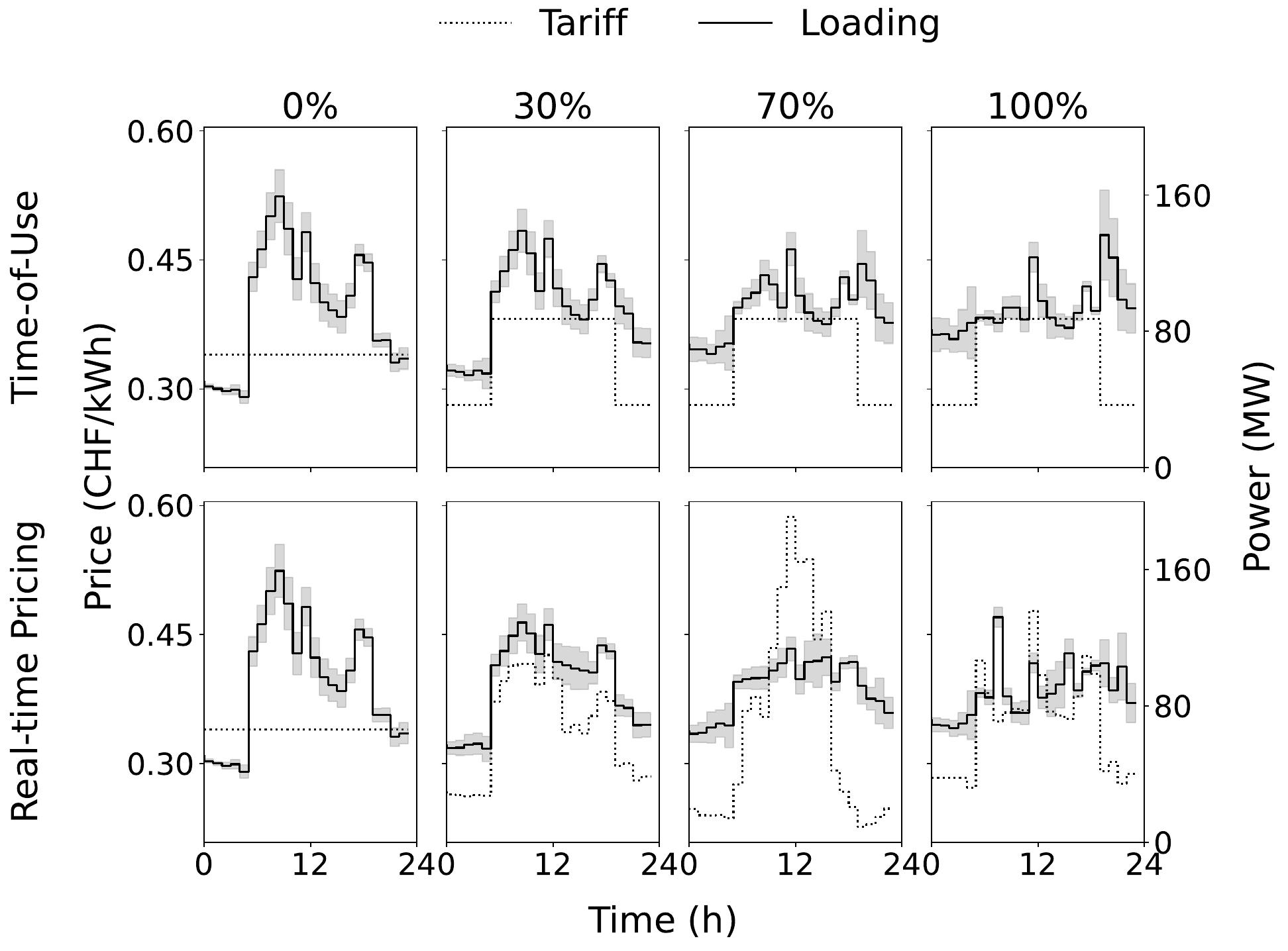}
    \captionof{figure}{Daily average and standard deviation bands for region-wide transformer loading over the simulated weeks, for selected tariff adoption levels (0\%, 30\%, 70\%, and 100\% plotted by column) and across the 24-hour period.}
    \label{fig:trafo_loading}
\end{center}

\begin{center}
    \captionsetup{hypcap=false}
    \includegraphics[width=0.48\textwidth]{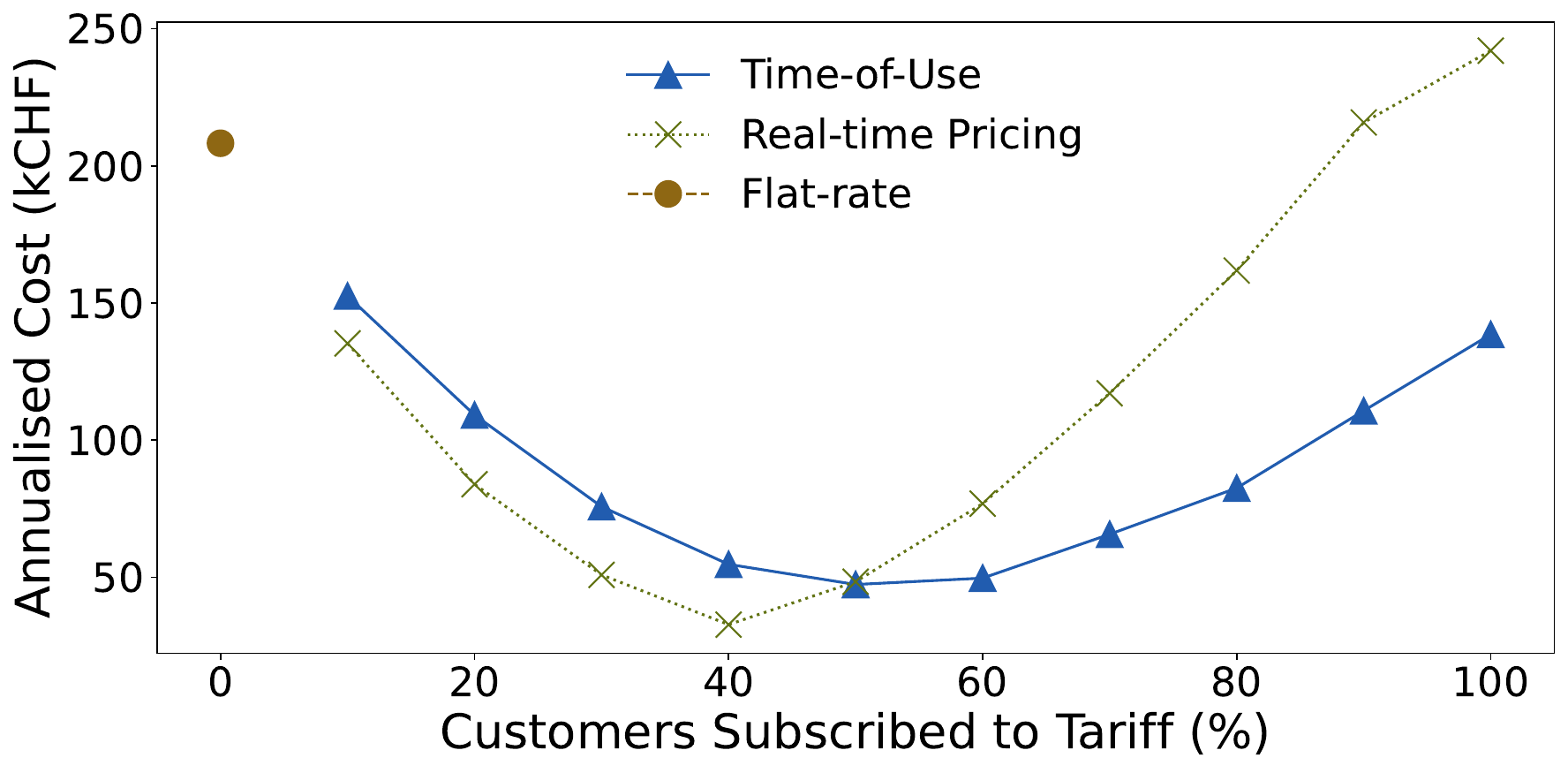}
    \captionof{figure}{Annualised reinforcement cost in the study region for Flat, \ac{ToU}, and \ac{RTP} tariffs at various levels of customer adoption.}
    \label{fig:expansion_cost}
\end{center}

\section{Results}
\label{sec:experiments}

The effects of time-varying tariff adoption on \ac{LVN} loading are analysed, as well as the expected network reinforcement costs.

\subsection{Operational Effects of Tariff Adoption}
\label{sec:op_effects}

Figure~\ref{fig:trafo_loading} shows the aggregated transformer loading for Bern, averaged across the generated scenarios, for a representative 24-hour period under the \ac{ToU} and \ac{RTP} tariffs at various levels of adoption. At 30\% and 70\% adoption, we observe a general flattening of the load profile, with a more substantial effect observed with the \ac{RTP} tariff. At 100\% \ac{ToU} adoption, a new rebound peak forms at the start of the off-peak price period, while the \ac{RTP} tariff causes a morning peak at a time with a relatively low electricity price.

\subsection{Effect of Tariffs in Distribution Network Expansion}
\label{sec:exp_effects}

Figure~\ref{fig:expansion_cost} shows the annualised cost of reinforcing the considered \acp{LVN} for varying levels of customer adoption of time-varying tariffs. The 0\% adoption rate indicates that all prosumers subscribe to the flat tariff. At only 10\% adoption, cost reductions of 29\% and 36\% are achieved for the \ac{ToU} and \ac{RTP}, respectively. Maximum reductions of 78\% and 85\% are achieved for the \ac{ToU} and \ac{RTP}, respectively, at adoption levels of 50\% and 40\%.

At 50\% and 40\% customer adoption for the \ac{ToU} and \ac{RTP}, respectively, there is a breakpoint after which the reinforcement costs increase with adoption. This increase is driven by demand synchronisation, as a higher share of demand responds to pricing signals. Notably, for \ac{RTP} adoption rates of 90\% or higher, reinforcement costs can exceed those required under a flat-rate subscription. 


\section{Conclusions}
\label{sec:conclusions}

This article proposes a \ac{MILP} model for the reinforcement planning of \acp{LVN} that considers prosumer adoption of time-varying tariffs. By conducting experiments on 471 reference \acp{LVN} in Bern, Switzerland, under projected \ac{DER} deployment in 2050, we demonstrate the impact of tariff adoption rates. The analysis reveals that at a 20\% adoption level, annualised reinforcement costs can decrease by up to 50\%, with \ac{RTP} being marginally more effective than \ac{ToU}. Nevertheless, we observe that reinforcement costs begin to rise at adoption rates above 50\% for \ac{ToU} and 40\% for \ac{RTP}. These results highlight the risk of prosumer synchronisation, as a critical mass of prosumers shift their load to low-price times. Furthermore, for adoption rates above 90\% of \ac{RTP}, the reinforcement needs can exceed those under flat-rate subscriptions.


\section*{Acknowledgements}
The research published in this publication was carried out with the support of the Swiss Federal Office of Energy SFOE as part of the SWEET EDGE and SWEET PATHFNDR projects. 

\bibliographystyle{elsarticle-num-names}
\bibliography{refs}

\end{multicols}
\end{document}